# Quantifying phonon particle and wave transport in nanostructures

## --The unexpectedly strong particle effect in silicon nanophononic metamaterial with cross junction

Dengke Ma[1,2,#], Anuj Arora[1,3,#], Shichen Deng[1,2], Junichiro Shiomi[3,4,*], and Nuo Yang[1,2,*]

[1]State Key Laboratory of Coal Combustion, Huazhong University of Science and Technology (HUST), Wuhan 430074, P. R. China

[2]Nano Interface Center for Energy(NICE), School of Energy and Power Engineering, Huazhong University of Science and Technology (HUST), Wuhan 430074, P. R. China

[3]Department of Mechanical Engineering, The University of Tokyo, 7-3-1 Hongo, Bunkyo, Tokyo 113-8656, Japan

[4]Center for Materials research by Information Integration, National Institute for Materials Science, 1-2-1 Sengen, Tsukuba, Ibaraki 305-0047, Japan

[#]D. M. and A. A. contributed equally to this work.

Electronic mail: J.S. (shiomi@photon.t.u-tokyo.ac.jp) and N.Y. (nuo@hust.edu.cn)

**ABSTRACT**

Understanding phonon transport mechanisms in nanostructures is of great importance for delicately tailoring thermal properties. Combining phonon particle and wave effects through different strategies, previous studies have obtained ultra-low thermal conductivity in nanostructures. However, phonon particle and wave effects are coupled together, that is their individual contributions to phonon transport cannot be figured out. Here, we present how to quantify the particle and wave effects on phonon transport by combining Monte Carlo and atomic green function methods. We apply it to 1D silicon nanophononic metamaterial with cross-junctions, where it has been thought that the wave effect was the main modulator to block phonon transport and the particle effect was negligibly weak. Surprisingly, we find that the particle effect is quite significant as well and can contribute as much as 39% to the total thermal conductivity reduction. Moreover, the particle effect does not decrease much as the cross section area (CSA) of the structure decreases and still keeps quite strong even for CSA as small as 2.23 nm$^2$. Further phonon transmission analysis by reducing the junction leg length also qualitatively demonstrates the strong particle effect. The results highlight the importance of mutually controlling particle and wave characteristics, and the methodologies for quantifying phonon particle and wave effect are important for phonon engineering by nanostructuring.

## Introduction

Controlling phonon transport to achieve unique properties is significantly important in various applications such as thermal management, thermal rectifier, and thermoelectric energy conversion.[1] A phonon in condensed matter is a quantized lattice vibration which exhibits both particle and wave nature.[1-3] Over the past decades, most of the commonly exercised approaches to manipulate phonon transport has focused on its particle nature. By introducing surface,[4-7] interface,[8-11] random pores[12,13] and dopants [14-18], phonons can be scattered and result in a reduced thermal conductivity.

Another line to manipulate phonon transport is based on its wave nature. With regard to the wave nature, the phase information of phonon must be considered, and the wave effects such as interference and resonance manifest within the length scales that the phase is coherent.[2,19,20] Such phonon coherent transport can lead to stop-band formation in periodic structures[21-25], local resonance hybridization[26-29], natural graded thermal conductivity,[30,31] and can be exploited for applications like phonon nano-capacitor[32,33]. To maintain the phase information of phonon, nanostructure boundaries need to be smooth enough to specularly reflect phonons.[2] In general, it is considered that the wave effect governs phonon transport at room temperature when the size of nanostructure goes to few nanometers.[3,26,27]

Combing phonon particle and wave transport can give rise to the novel minimum thermal conductivity in superlattice,[3,24] and can significantly reduce thermal conductivity of nanostructures.[27] However, in presence of both phonon particle and wave effects, their individual contribution to the modulation of thermal properties has not been understood. Since these two effects are governed by different physical laws and can be engineered through different strategies,[1,2,26,34] quantitative understanding of the particle and wave effect on phonon transport will be of great importance to tune the thermal properties.

The nanophononic metamaterial (NPM) is based on the phonon resonance hybridization wave effect, [26-28] which generally consists of junction systems.[26-28] In the junction part, the resonant modes hybridize with the propagating modes and reduce their group velocity. The silicon nanowire cross junction (NCJ) is a typical NPM.[26-28] Using NCJ either in a form of either 2D network or 3D cage, the band modulation due to wave effect does not require the structure to be periodic since the resonance is local, unlike the phonon crystals[21,22,25]. This simplifies the synthesis of NCJ-based NPM, making it a good candidate for thermoelectrics. Previous studies about the resonance hybridization in NPM have mainly focused on phonon wave effect, considering the particle effect to be negligibly weak.[26,27]

In this paper, we propose to quantify the balance between the phonon particle and wave transport by combining Monte Carlo (MC) and atomic green function (AGF) method. We first briefly discuss our methodology, and then carry out a benchmark study on silicon-nanowire-cage to demonstrate the accuracy of our methods. We take NCJ structure as an example,[28] and quantify the contributions of phonon particle and wave effects on the reduction of thermal conductivity with varying cross section area ($CSA_W$) and leg numbers. Phonon transmission studies, by changing the leg length, are also carried out to support the quantitative results.

## Methodology

We take NCJ as an example, and explain the framework of our simulation method. Figure 1 (a) shows the simulation cell of NCJ, consisting of wires that are connected by cross junction. Despite previous studies demonstrating that there is a strong wave effect in the junction part,[26-28] in order to probe the phonon transport from partible standpoint, we use MC method [35-37], which solves Boltzmann transport equation based on phonon particle transport. Therefore, the MC simulation will ignore the wave effect and hence, surely overestimate the thermal conductivity.

To accurately simulate phonon transport in NCJ, wave effect must be included. As the cross junction effect is local,[26-28] the entire structure of NCJ can be divided into wire part and junction part (as shown in Fig. 1 (a)). When phonons propagate in the wire part, they undergo phonon-phonon scattering and boundaries scattering. So, the purely particle based MC can be used to describe the phonon transport in the wire part. When phonons reach the junction part, wave effect becomes relevant and hence needs to be considered. Here, we treat the whole junction part as an interface (as shown in Fig. 1 (a)), and use the transmittance, which describes the fraction of the incident phonons of frequency $\omega$ transmitting through specific area, to decide whether the phonons can transport across the junction part (interface). The phonon wave information is included in the transmittance ($\tau(\omega)$).

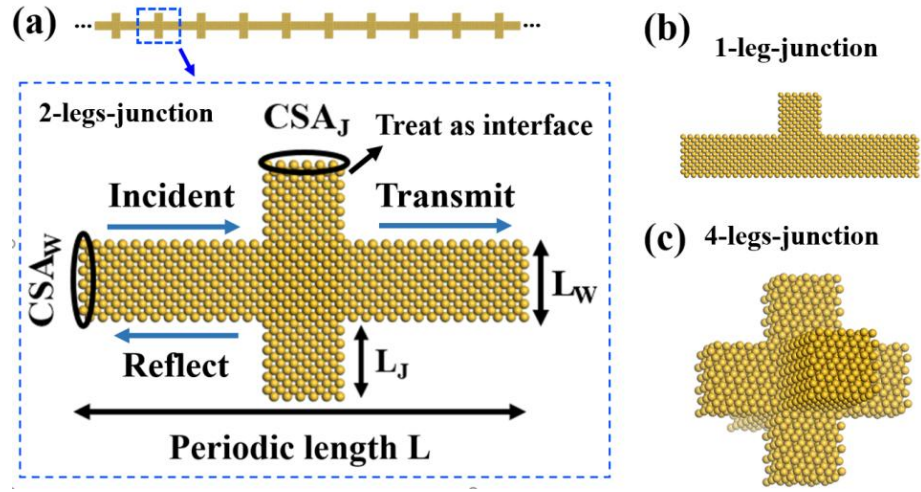

Fig.1 (a) Schematic picture of the phonon transport across NCJ. The structure of (a) 2-legs-junction, (b) 1-leg-junction and (c) 4-legs-junction. The length of the junction wire $L_J$ is $0.25a$ larger than the side length of main wire $L_W$, where $a$ is the lattice parameter of silicon and equals to 0.5431 nm. The CSA of the junction wire $CSA_J$ is the same as the CSA of main wire $CSA_W$. L is the periodic length of the total structure, which is 4 times the side length of main wire $L_W$.

To incorporate the transmittance into MC, a random number ($r$) is drawn from a uniform distribution ($r \in [0, 1]$) for every phonon, approaching the interface. If $r < \tau(\omega)$, the phonons will propagate across the junction part (interface). These phonons can then contribute to thermal conductivity. For the transmitted phonons, the velocity is kept the same. If $r \geq \tau(\omega)$, the phonons undergo specular reflection. In this case, the velocity is reassigned as Eq. (1)

$$\vec{V}_r = \vec{V}_i + 2|\vec{V}_i \cdot \vec{n}|\vec{n} \qquad (1)$$

where $\vec{V}_r$ and $\vec{V}_i$ is the reflected and incident phonon velocity, respectively. $\vec{n}$ is the unit normal to the interface.

The AGF method [38-40] is employed to get $\tau(\omega)$. Firstly the phonon transmission function $\Xi(\omega)$ is calculated.(Simulation details of AGF are provided in SII) The physical meaning of the transmission function is the number of phonon modes transmitted through the center device at a specific frequency.[39] Then the transmittance can be related to transmission function as[38]

$$\tau(\omega) = \frac{\Xi(\omega)}{\Xi_1(\omega)} \tag{2}$$

where $\Xi_1(\omega)$ is the ideal transmission function for the case that all phonon modes are transmitted without being scattered or reflected back from the central device to the heat bath (shown in Fig. S1(a)). $\Xi(\omega)$ is the transmission function of phonons through the junction part(Fig. S1(a)). When calculating $\Xi_1(\omega)$ and $\Xi(\omega)$, the heat bathes are the same semi-infinite long wires. Therefore, the transmittance $\tau(\omega)$ describes the fraction of the incident phonons of frequency $\omega$ that are transmitted through a center device. Consequently, its value lies between zero and unity. (Fig. S1(b)) In the simulation, to get $\Xi_1(\omega)$, we set the center region as wire with the length along the wire direction to be 0.543 nm. The thermal transport through such a short wire is ballistic, and as expected, the transmission function is practically equal to the total number of phonon modes (shown in Fig. S1(a)).

By using the phonon transmittance by AGF to incorporate phonon wave effect to MC, both phonon particle and wave effect are included. In the following manuscript, we term this method AGFMC.

To quantify phonon particle and wave effect in cross junction, we use MC to calculate the thermal conductivity of silicon nanowire (SiNW) ($\kappa_{SiNW}$) and NCJ ($\kappa_{NCJ\_MC}$) separately. Then, we should use AGFMC to calculate the thermal conductivity of the same NCJ ($\kappa_{NCJ\_AGFMC}$). The CSA and periodic length of SiNW are the same as the $CSA_W$ and L of NCJ (as shown in Fig.1(a)), respectively. As MC is based on phonon particle transport, the difference between $\kappa_{SiNW}$ and $\kappa_{NCJ\_MC}$ is solely due to particle effect of the cross junction. On the other hand, the difference between $\kappa_{NCJ\_MC}$ and $\kappa_{NCJ\_AGFMC}$ is due to the wave effect of the cross junction.

To quantitatively show the phonon particle and wave effects, we define a parameter $\eta_{wave}$, which measures the fraction of thermal conductivity reduction by wave effect to the total thermal conductivity reduction as,

$$\eta_{wave} = \frac{\kappa_{NCJ\_MC} - \kappa_{NCJ\_AGFMC}}{\kappa_{SiNW} - \kappa_{NCJ\_AGFMC}} \tag{3}$$

$$\eta_{particle} = 1 - \eta_{wave} \tag{4}$$

## Results and discussions

In order to validate the proposed AGFMC method, we consider the silicon nanowire cage structure (SiNWC).[31] The simulation cell for the SiNWC is NCJ (as shown in Ref. 28). We first calculate the thermal conductivity of SiNW for different cross section area through MC. (simulation details of MC are provided in SII) As shown in Fig. S3, the computed $\kappa_{SiNW}$ agrees well with previous MC and analytical results, validating our MC simulations. Moving on, we simulate the SiNWC in Ref. 28. We use both MC and AGFMC methods to calculate the thermal conductivity of SiNWC ($\kappa_{SiNWC}$), respectively. As shown in Fig. 2, the $\kappa_{SiNWC}$ (black dot) calculated by MC is significantly larger than the previous molecular dynamicals (MD) result[28] (blue dot). On the other hand, $\kappa_{SiNWC}$ by AGFMC (red dot) is close to the MD result, and much smaller than the MC result. This means that by treating phonons simply as particle, MC vastly overestimates the $\kappa_{SiNWC}$ because it misses the strong wave effect in SiNWC.[26,28] On the contrary, by incorporating wave effect, the AGFMC can get a much better prediction for $\kappa_{SiNWC}$. There are some discrepancy between the AGFMC and MD values but this is reasonable considering that the difference in the methods, and the potential used in the MD is different from that used to calculate the phonon properties in MC. In addition, the AGF does not include high order phonon process,[38,39] and the specularity parameter in MC is an adjustable parameter. In fact, this difference is one of the reasons why we chose to incorporate wave effect in the AGFMC method rather than taking the MD results in what follows aiming to quantify the phonon wave effect by comparing with MC simulation.

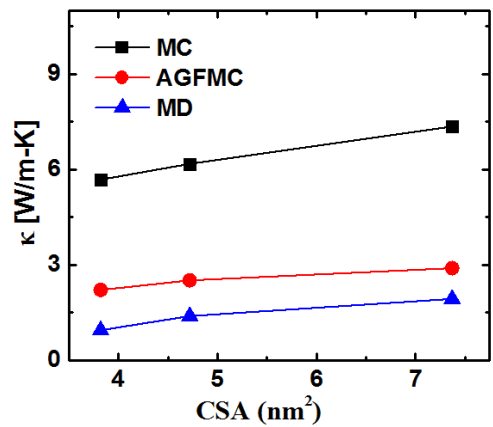

Fig.2 The thermal conductivity of silicon nanowire cage versus the CSA at 300 K. The data of black are obtained by the MC method, which only takes phonon particle effect into account. The data of blue are obtained by the AGFMC method, which accounts for both phonon particle and wave effects.

Now we focus on the NCJ structure. Periodic boundary condition is applied along the wire direction, and free boundary condition is applied perpendicular to the wire direction (net heat flux direction). In our simulation, as shown in Fig.1(b), the length of the junction wire $L_J$ is $0.25a$ larger than the side length of the main wire $L_W$, where $a$ is the lattice parameter of silicon and equals to 0.5431 nm. And the CSA of the junction wire $CSA_J$ is the same as the CSA of main wire $CSA_W$. L is the periodic length of the total structure, which is 4 times the side length of main wire $L_J$. Thus, as the $CSA_W$ increases, the NCJ structure effectively scales up.

Figure 3(a) shows the thermal conductivity of 2-leg-junction (as shown in Fig. 1(b)). $\kappa_{SiNW}$, $\kappa_{NCJ\_MC}$ and $\kappa_{NCJ\_AGFMC}$ all increase when the CSA ($CSA_W$) increases from 2.23 $nm^2$ to 17.72 $nm^2$. This is because for SiNW, the phonon transport is mainly governed by phonon boundary scattering.[4] As the CSA increases, the surface-to-volume ratio decreases, and the boundary scattering becomes weaker. For NCJ, when considering only particle effect ($\kappa_{NCJ\_MC}$), besides the boundary scattering in the wire part, the phonon scattering in the junction part also becomes weaker with the size of junction increasing. When both particle and wave effect are taken into account ($\kappa_{NCJ\_AGFMC}$), besides the boundary scattering, the wave effect also becomes weaker. These mean that the phonons are more likely to transport across the junction part. As shown in Fig. 4(a), the AGF calculated transmittance increases as the CSA increases for the whole frequency range.

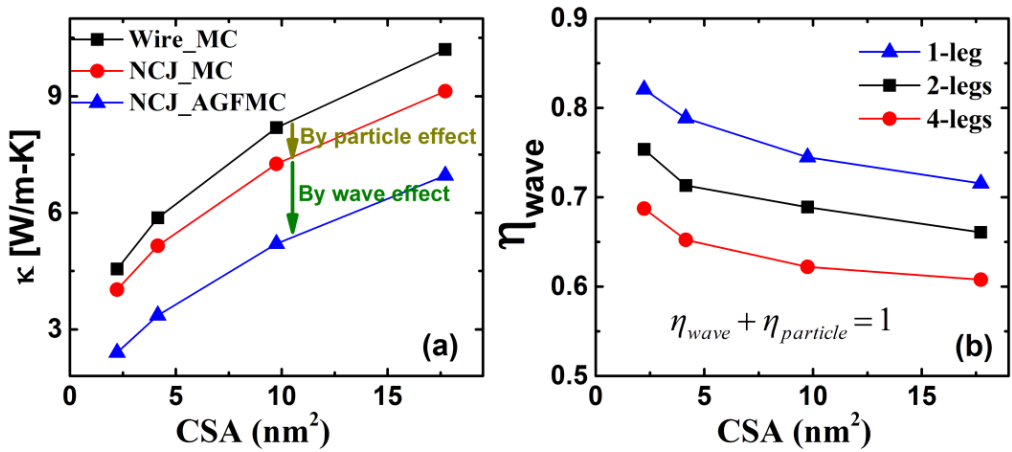

Fig.3 (a) The thermal conductivity of SiNW (black dot) and NCJ (red and blue dot) versus the CSA at 300 K. The data of red line are obtained by the MC method, which only takes phonon particle effect into account. The data of blue line are obtained by the AGFMC method, which accounts for both phonon particle and wave effects. (b) The ratio of thermal conductivity reduction by phonon wave effect to the total thermal conductivity reduction versus the CSA. Both the CSA referred here to $CSA_W$ (as shown in Fig. 1(b)).

More importantly, with the introduction of the cross junction, $\kappa_{NCJ\_MC}$ (red dot) are smaller than $\kappa_{SiNW}$ (black dot). This is because the cross junction increases phonon scattering. In addition, $\kappa_{NCJ\_AGFMC}$ (blue dot) are even smaller than $\kappa_{NCJ\_MC}$ (red dot). This, as discussed earlier, is due to the enhanced blockage originating from phonon resonance hybridization wave effect, which has been incorporated into AGFMC.[26-28] Using Eq. (3), we calculate the $\eta_{wave}$ of 1-leg-junction, 2-legs-junction and 4-legs-junction with the varying of CSA ($CSA_W$) (as shown in Fig. 1(c) and 1(d)). When the CSA increases from 2.23 nm$^2$ to 17.72 nm$^2$, the $\eta_{wave}$ decreases monotonously for 1-leg-junction, 2-legs-junction and 4-legs-junction. This shows that the wave effect weakens with the system size increasing, which is in accordance to the previous studies.[3,24,41] When the system size gets large enough, the wave effect due to resonance hybridizations will diminish,[41] and the phonon transport turns to incoherent.[3]

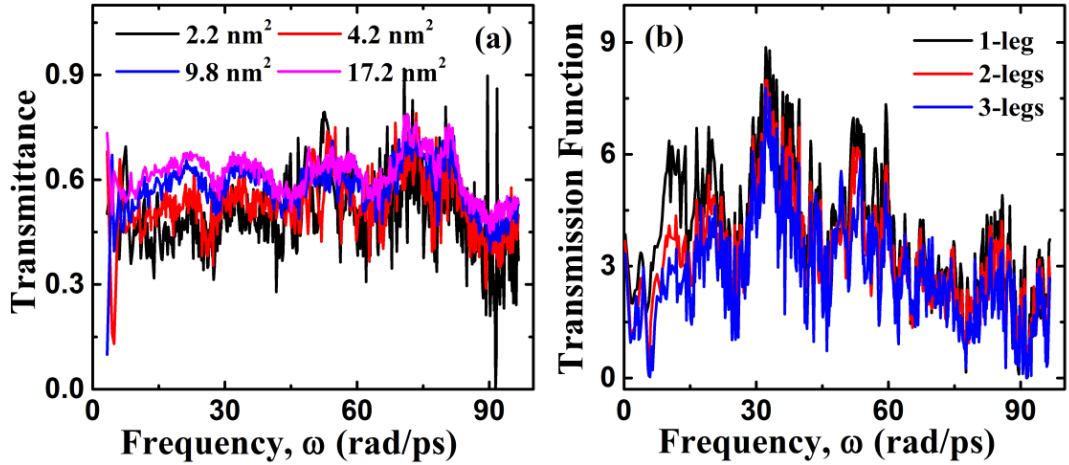

Fig.4 (a) Frequency dependent transmittance of 4-legs junction of NCJ with different CSA. (b) Frequency dependent transmission function of 1-leg NCJ (black line), 2-legs NCJ (red line) and 4-legs NCJ (blue line) with leg length equals to 3$a$, where a is the lattice parameter of silicon that equals to 0.5431 nm.

What's more striking is that the phonon particle effect has a significant contribution to thermal conductivity reduction, and cannot be neglected. As shown in Fig. 3(b), when the CSA increases from 2.23 nm$^2$ to 17.72 nm$^2$, the $\eta_{wave}$ for 1-leg-junction decreases from 0.82 to 0.72, and the $\eta_{wave}$ for 4-legs-junction decreases from 0.69 to 0.61. Conversely, $\eta_{particle}$ increases from 0.18 to 0.39 (shown in Fig. S6(b)). With the CSA increases, the $\eta_{wave}$ will surely increase. Furthermore, what needs to be emphasized is that with the CSA decreases, the $\eta_{particle}$ does not decrease much (shown in Fig. S6(b)). The $\eta_{particle}$ is quite large even for CSA as small as 2.23 nm$^2$, which accentuates the importance of mutually control of the particle and wave characteristics in NPM.

To have a deep understanding of the strong particle effect in NCJ, we calculate the transmission function with the decreasing of leg length $L_J$ of NCJ. We focus on 4-legs-junction with CSA$_W$ and CSA$_J$ equal to 2.23 nm$^2$. As shown in Fig. 5 (a), when the leg length decreases from 3$a$ to 0.25$a$, the transmission function increases, which implies that the junction effect becomes weaker. This tendency is in accordance to previous result.[27,41] While as shown in Fig. 5(b), when L$_J$ is reduced to just one atomic layer (0.25$a$), the transmission function of 4-legs-junction (black line) is still quite lower than the corresponding SiNW (green line). This means that the leg, even with just one atomic layer, still has a strong impact on the phonon transport. To further validate this effect, we change the 4-legs-juntion to 1-leg-juntion keeping $L_J$=0.25$a$. Comparing with the corresponding 4-legs-junction, the transmission for 1-leg-junction (red line) increases, but still is much smaller than that for SiNW (green line). As the resonance hybridization wave effect is the resonance modes in the leg part that interacts with the propagating modes in the main wire part.[26,27] When the leg length $L_J$ equals to merely one atomic layer, the wave resonance hybridization effect on the phonon transport becomes very weak.[26,42] The reduction of transmission function from the SiNW to the 1-leg-junction with $L_J$ corresponding to one atomic layer, is induced mainly by scattering. And the big gap between these two indicates that the scattering is quite strong.

Besides, as the leg length decreases, the transmission function, for NCJ, at low frequencies gradually approaches that of SiNW. While the transmission function, for NCJ, at high frequencies is still quite different from that of SiNW. This is because the wave resonance hybridization can block low frequency phonons.[26,27] But the scattering typically has a strong effect on high frequency phonons.[27] Therefore, for low frequency phonons, when the leg length decreases to just one atomic layer, the resonance hybridization effect becomes very weak. The transmission function of low frequency phonon for NCJ definitely approaches to SiNW. While the high frequency phonons of NCJ still suffer scattering, their transmission function is still lower than that of SiNW.

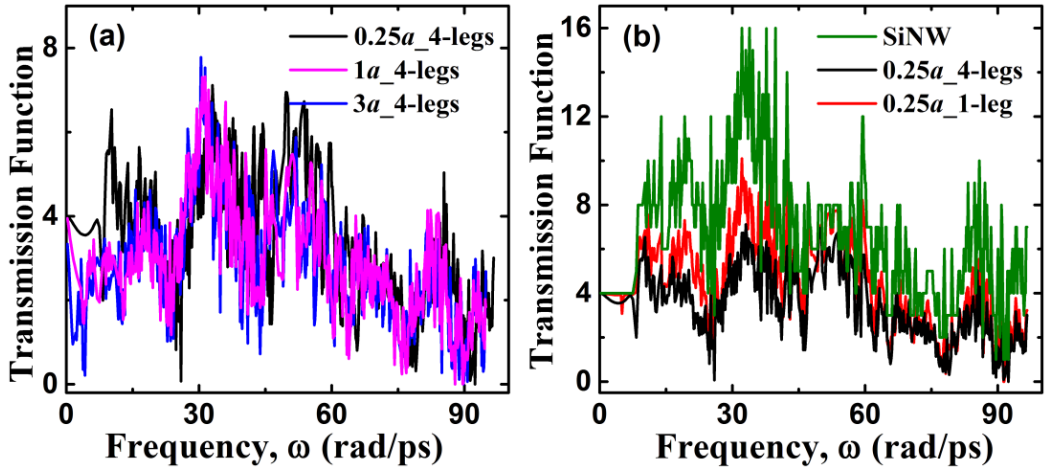

Fig.5 Frequency dependent transmission function of SiNW (green line), 1-leg NCJ (red line), and 4-leg NCJ. 3$a$, 1$a$ and 0.25$a$ are the leg length. And $a$ = 0.5431 nm is the lattice parameter of silicon, and thus, 0.25$a$ corresponds to length of single atomic layer.

We should also notice that although the increasing of junction numbers will further blocks phonon transport and reduces the transmission function (as shown in Fig. 4(b)) and the $\kappa_{NCJ\_AGFMC}$ (as shown in Fig. S4). [27,28] The $\eta_{wave}$ decreases when the structure changes from 1-leg-junction to 4-legs-junction. Because more junctions will induce both more resonance hybridizations[27] and phonon scattering (as shown in Fig. S5).[26,28] Since, the junctions are located on different sides of the main wire, phonon scattering induced by junctions at different sides of the main wire will slightly correlate with each other. While for resonance hybridization, although the number of resonant phonons increases with the increasing in junction number, the number of propagating phonons remains the same. The resonant phonons eventually should interact with the propagating phonons in the main wire and correlate with each other, thus, leads to the decreasing of $\eta_{wave}$.

In conclusion, we incorporate phonon wave effect to MC by AGF, and propose to use the AGFMC method to simulate phonon transport in nanostructure where the phonon wave effect is important. The AGFMC has the advantage to handle bulk random nanostructures. The Benchmark studies on SiNWC validates the accuracy of AGFMC method. By performing simulation on SiNW and NCJ with MC and AGFMC, we quantify the phonon particle and wave effect in NCJ based NPM. We find that an increasing system size leads to a weakening of wave effect. Though an increasing number of junctions definitely increase the blockage to phonon transport, the relative portion of wave effect decreases. Interestingly, we find that the particle effect, which is thought to be negligibly weak previously in resonance hybridized NPM, can actually contribute a lot to the thermal conductivity reduction. The $\eta_{particle}$ is quite large even when the CSA is as small as 2.23 nm$^2$, and can be as high as 0.39 for 4-legs-junction NCJ when the CSA (CSA$_W$) is 17.72 nm$^2$. The big gap of transmission function between the SiNW and 1-leg-junction with $L_J$ corresponding to one atomic layer confirms the strong particle effect qualitatively. The results here accentuate the importance of mutually controlling particle and wave characteristics. The methodologies to quantitatively identify phonon particle and wave effect will further pave the way for phonon engineering.

## Acknowledgments

N.Y. is sponsored by National Natural Science Foundation of China (No. 51576076 and No. 51711540031), Hubei Provincial Natural Science Foundation of China (2017CFA046) and Fundamental Research Funds for the Central Universities (2016YXZD006). J.S. is supported in part by Bilateral Joint Research Project and KAKENHI Grants No. 16H04274 from Japan Society for the Promotion of Science (JSPS), "Materials research by Information Integration" Initiative (MI2I) project, and CREST Grant No. JPMJCR16Q5 from Japan Science and Technology Agency (JST). The authors are grateful to Lei Feng, Masato Onishi and Donguk Suh for fruitful discussions. The authors thank the National Supercomputing Center in Tianjin (NSCC-TJ) and China Scientific Computing Grid (ScGrid) for providing assistance in computations.

**Competing financial interests**: The authors declare no competing financial interests.